\documentstyle[prd,aps,floats,psfig]{revtex}
\begin{document}
%\preprint{KNGU-INFO-PH-22, BROWN-HET-, hep-ph/0207246}
\draft

\renewcommand{\topfraction}{0.99}
\renewcommand{\bottomfraction}{0.99}
\twocolumn[\hsize\textwidth\columnwidth\hsize\csname
@twocolumnfalse\endcsname

\title{\Large Stabilization of the Electroweak Z String in the Early
Universe}

\author{ M. Nagasawa$^{1,2}$ and R. Brandenberger$^2$}
\address{~\\$^1$Department of Information Science, Faculty of Science,
Kanagawa University, Kanagawa 259-1293, JAPAN; 
~\\$^2$Physics Department, Brown University, Providence, RI 02912,
USA.}
\date{\today}
\maketitle

\begin{abstract}
The standard electroweak theory admits a string solution, the Z string, in
which only the electrically neutral Higgs fields are excited. This solution
is unstable at zero temperature: Z strings decay by exciting charged Higgs
modes. In the early Universe, however, there was a long period during which
the Higgs particles were out of equilibrium but the photon field was in
thermal equilibrium. We show that in this phase Z strings are stabilized by
interactions of the charged Higgs modes with the photons. 
In a first temperature range immediately below the electroweak symmetry
breaking scale, the stabilized embedded defects are symmetric in
internal space (the charged scalar fields are not excited). 
There is a second critical temperature below which
the stabilized embedded strings undergo a core phase transition and the
charged scalar fields take on a nonvanishing value in the core of the
strings. We show that stabilized embedded defects with an asymmetric 
core persist to very low temperatures. The stabilization
mechanism discussed in this paper is a prototypical example of a
process which will apply to a wider class of embedded defects in
gauge theories.
\end{abstract}

]

{\bf 1. Introduction}

In a previous paper \cite{NB99} we suggested that a wide class of embedded
defects may be stabilized by plasma effects. We studied a toy model
consisting of four real scalar fields $\phi_i \,\, i = 1, .., 4$ with the
``generalized Mexican hat" potential
\begin{equation} \label{toy}
V(\phi) \, = \, \lambda (\sum_{i=1}^4 \phi_i^2 - \eta^2)^2 \, ,
\end{equation}
two of which ($\phi_1$ and $\phi_2$ to be specific) being electrically
charged, the other two neutral. At zero temperature, the vacuum manifold in
this model is ${\cal S}^3$ and hence does not admit any stable topological
defects. However, there are {\it embedded} defects \cite{embedded,AV99},
configurations which span only a subset of the vacuum manifold. One example
is the global cosmic string solution of the subspace of the theory with
$\phi_1 = \phi_2 = 0$. At zero temperature this string configuration is
unstable and can decay through the excitation of the charged fields. In more
graphic terms, the field configuration
slips off the top of the potential in the charged field directions and
approaches the vacuum manifold everywhere in space. However, in the presence
of a bath of photons, interactions of the photons with the charged scalar
fields will lead to an effective potential which is lifted in the directions
of the charged fields, thus generating a potential barrier which can
stabilize the embedded string. This toy model is a theory with a global
symmetry, and is realized in the Sigma model description of the
low energy limit of Quantum Chromodyanamics in the limit of vanishing
pion mass (see \cite{Carter:2002te} which includes a detailed discussion
of effects which arise when the explicit symmetry breaking caused by the
nonvanishing pion mass is taken into account).

The standard electroweak theory is a good example of a theory which contains
embedded strings in which only the neutral scalar field
components of the multi-dimensional order parameter are
excited. Among all possible embedded
strings, (see e.g. \cite{TV} for a classification of embedded strings in the
electroweak theory) the electroweak Z string \cite{Nambu,TV92} is the string
configuration consisting of excitations of the neutral fields only.

In the paper we demonstrate that Z strings are stabilized during
the temperature interval between the electroweak phase transition
temperature and the recombination. At a certain temperature during
this period, a core phase transition takes place \cite{CPT} below
which the embedded strings will become superconducting as shown in
\cite{Carter:2002te}, thus enhancing their stability
and leading to the formation of vortons \cite{vorton}. 
The enhanced stability of electroweak strings due to current-carrying
zero modes is similar to the stabilization by neutrino zero modes
considered in \cite{StaSto}.

The mechanism discussed in this paper generalizes to a wide class of
embedded defects. Since defects can play an important role in cosmology -
even if they
are not stable at all times - the mechanism discussed here may have
important consequences for many aspects of cosmology. One possibility is to
use embedded global anomalous strings such as the pion string of low energy
QCD \cite{pion} to generate primordial magnetic fields \cite{BZ98}. Crucial
aspects in this application are, besides the stabilization of the embedded
strings, the fact that, via the anomaly, charged zero modes on the string
generate coherent magnetic fields circling the string \cite{KM88}, and the
fact that the length scale of the string network increases in comoving
coordinates (see e.g. \cite{stringrev}), which provides the mechanism for
generating the required large coherence length of the primordial magnetic
fields. QCD at large baryon density also leads to the stabilization of a type
of embedded strings, K-strings \cite{Buckley:2002ur}.

Stabilized embedded defects (in particular embedded walls) could also play
a role in defect-mediated baryogenesis (see e.g. \cite{BDH} and
\cite{BD,BDPT} for the main ideas of defect mediated GUT and electroweak
baryogenesis, respectively), and they may be useful in implementing the
scenario of Dvali et al. \cite{DLT} for alleviating the monopole problem via
defect interactions (see \cite{ABES,PV} for studies of the basic interaction
mechanism). 

As with any class of topological defects, there will be
severe cosmological constraints on models which admit them, resulting
from the fact that the evolution of the defects in the early
Universe  may lead to predictions which are in conflict with 
observations. The
vorton abundance problem (see e.g. \cite{BCDT}) leads to severe 
constraints, as does the constraint (specific to decaying defects) that
the decay not lead to spectral distortions in the cosmic microwave
background \cite{BDC}.

A new feature which arises in the present discussion of the electroweak
theory, compared to the previous studies of toy models with a global
symmetry, is the fact that the underlying symmetry is a local symmetry.

{\bf 2. Stabilization Mechanism}

Starting point is the Lagrangian for the standard electroweak theory:
\begin{eqnarray}
{\cal L} \, &=& \, -{1 \over 4} W_{\mu \nu a} W^{\mu \nu a} - {1 \over 4}
Y_{\mu \nu} Y^{\mu \nu} \nonumber \\
&+& \left|(\partial_{\kappa} - {1 \over 2}i g \tau^a W_{\kappa}^a
- {1 \over 2} i g' Y_{\kappa}) \Phi\right|^2 \nonumber \\
&-& \lambda (\Phi^{\dagger} \Phi - {\eta^2})^2 \, , \label{stdlag}
\end{eqnarray}
where the indices $\mu$ and $\nu$ are Lorentz indices, the index $a$ is an
SU(2) index, $g$ and $g'$ are the SU(2) and U(1) coupling constant,
respectively, and
$W$ and $Y$ are the SU(2) and U(1) field strength tensors, respectively.

The field $\Phi$ is a complex Higgs doublet, the upper component $\Phi^+$
having positive charge, the lower component $\Phi^0$ being neutral
(if the expectation value of $\Phi$ is chosen such that only the
lower component is non-vanishing). The
analogy with (\ref{toy}) is clear: the lower SU(2) doublet component gives
the two neutral real scalar fields $\phi_3$ and $\phi_4$, the upper charged
component yields $\phi_1$ and $\phi_2$.

We are interested in describing the physics below the phase transition
temperature $T_c$. In this case, the only gauge field which is excited is
the photon field $A_{\mu}$. In terms of the SU(2) gauge field $W^a$ and the
U(1) gauge field $Y$ (we are suppressing the Lorentz index), the $Z$ and $A$
fields are given by
\begin{eqnarray}
Z \, &=& \, \cos(\theta_w) W^3 - \sin(\theta_w) Y \nonumber \\
A \, &=& \, \sin(\theta_w) W^3 + \cos(\theta_w) Y \, , \label{trans}
\end{eqnarray}
where $\theta_w$ is the weak mixing angle.

The electroweak Z string is obtained by setting
\begin{eqnarray}
W^1 = W^2 = A = \Phi^+ &=& 0 \nonumber \\
\Phi^0 &=& \Phi_{NO} \\
Z &=& A_{NO} \nonumber \, ,
\end{eqnarray}
where the subscript {\it NO} stands for the Nielsen-Olesen U(1) cosmic
string configuration \cite{NO}, which in cylindrical coordinates $r$ and
$\theta$ (the coordinates in the plane perpendicular to the string) and in
temporal gauge $A_{\mu = 0}^i = 0$ has the form
\begin{eqnarray}
\Phi_{NO}(r, \theta) &=& \eta f(r) e^{i \theta} \nonumber
\\
A_{\mu, NO}(r, \theta) &=& -{{v(r)} \over {\alpha r}} \delta_{\mu \theta} \,
,
\end{eqnarray}
where $f(r)$ and $v(r)$ are the profile functions of the string.
Note that since for the electroweak Z-string the vacuum expectation
value of the upper component of $\Phi$ always vanishes, the upper
component of $\Phi$ will be associated with electrically charged
degrees of freedom everywhere in space.

At this point the reader may object and recall that in the electroweak
theory there is only one physical Higgs degree of freedom. At any
point in space one can choose a gauge in which only one of the neutral
Higgs fields is non-vanishing. The three scalar field degrees of freedom
corresponding to rotations in the vacuum manifold are eaten by the
gauge fields acquiring masses, and it is only the massive Higgs degree
of freedom which remains. However, the existence of topological defects
in gauge theories is precisely a consequence of the fact that there are
restrictions on the ability to impose the abovementioned gauge uniformly
over space. By causality, there will be regions in space in which the
Higgs field is not in its vacuum manifold. This corresponds to
localized energy configurations which cannot be gauged away.
To describe the physics of such defects, it is advantageous to use
a gauge in which the Higgs fields are not fixed, as we do below.

In the absence of any excitations in the bulk, the electroweak Z string is
unstable \cite{instab}. The unstable mode corresponds to an excitation of
the charged Higgs doublet:
\begin{eqnarray}
\Phi(\xi, {\vec x}) &=& \cos(\xi) \Phi^0(\cos(\xi) {\vec x}) + \sin(\xi) \Phi^+
\nonumber \\
Z^j(\xi, {\vec x}) &=& \cos(\xi) Z_j^{(0)}(\cos(\xi) {\vec x}) \, ,
\label{unst}
\end{eqnarray}
where $\xi$ is the deformation parameter (for $\xi = 0$ the configuration
reduces to the Z string), $Z_j^{(0)}$ is the Nielsen-Olesen gauge field
configuration, and $\Phi^0$ is a complex Higgs doublet whose neutral
component takes on the Nielsen-Olesen string configuration, and whose
charged component vanishes. The field $\Phi^+$ stand for the complex
Higgs doublet in which only the charged complex scalar
is excited, and its transpose of $\Phi^+$ is given
by
\begin{equation}
(\Phi^+)^T \, = \, \eta (1, 0) \, .
\end{equation}

Since by escaping into the charged scalar field directions, the string
configuration can decrease its potential energy in the string core region,
the energy per unit length $E(\xi)$ of the configuration (\ref{unst}) is
smaller than the energy per unit length $E_0$ of the Z string. Since the
loss in energy density is proportional to $\xi^2 V(0)$, and since the loss
in energy density is confined to the core region of the string (core radius
$r_c$), we obtain the following estimate for the maximal energy loss:
\begin{equation}
E(\xi) \, \geq \, E_0 - \kappa \lambda \xi^2 \eta^4 r_c^2 \, ,
\end{equation}
where $\kappa$ is a constant of order 1. A detailed analysis \cite{instab}
shows that for realistic values of the weak mixing angle, there is indeed an
energy loss.

Another way to view the instability of the electroweak string at zero
temperature \cite{Perkins} is as an instability to the formation of a
W-condensate in the core of the string, a process which was shown to
lower the energy. In \cite{AA} it was shown that this instability
can be gauge transformed to an instability lowering the winding
number $N$ of the pure electroweak string by 1 (at least for $N > 1$).
Thus, this instability is not independent of the one analyzed
in \cite{instab}.
Since in our numerical work we focus on the Higgs sector, we will also
focus our analytical considerations mostly on the instability mode
discussed above (\ref{unst}), although we will estimate the energy
gain and loss by this process at the end of this section.

We will now show that in the presence of a background thermal bath of
photons, the Z string is stabilized.
For our analysis to hold, it is important that all of the scalar fields be
out of thermal equilibrium, that the photon is in thermal equilibrium, and
that the other gauge fields be out of equilibrium. These conditions are
naturally satisfied in the electroweak theory at temperatures below the
electroweak symmetry breaking scale and below the mass of the Higgs
particle, but above the temperature of recombination. In this (large)
temperature interval it is justified to average over the light degrees of
freedom in order to study the dynamics of the order parameter.

Thus, our procedure will be to consider the Lagrangian of the electroweak
theory in the presence of a thermal bath of photons. We take the thermal
average of this Lagrangian and extract the terms which act as a correction
to the potential for the dynamics of the order parameter. The thermal
averaging consists of
setting averages of $A_{\mu}$ to zero, and making the replacement
\begin{equation} \label{av}
A_{\mu} A^{\mu} \, \rightarrow \, - \alpha T^2 \, ,
\end{equation}
where $\alpha$ is another positive constant of order 1 (the
estimate of \cite{Carter:2002te} gives a value somewhat larger
than 1, and the larger the number of degrees of freedom in
thermal equilibrium in the plasma, the larger the value of
$\alpha$ will be).

Thus, the starting point is the Lagrangian (\ref{stdlag}). In this
Lagrangian, we set the charged gauge fields to zero, and invert the
transformation (\ref{trans}) of the neutral fields in order to express the
Lagrangian in terms of $A_{\mu}$ and $Z_{\mu}$. If the charged scalar fields
are excited as in (\ref{unst}), then the transformation (\ref{trans})
is effected. However, the effect will be a correction of order $\xi^2$
in a quantity which is already of order $\xi^2$ and can hence be neglected.
Thus, we use
\begin{eqnarray}
W^3 \, &=& \, \cos(\theta_w) Z + \sin(\theta_w) A \nonumber \\
Y \, &=& \, \cos(\theta_w) A - \sin(\theta_w) Z \, .
\end{eqnarray}
Inserting these transformations into (\ref{stdlag}) allows us to extract the
extra contribution to the potential energy density of the scalar fields
which stems from the presence of the bath of photons. This contribution
comes from the part of the covariant derivative term which is quadratic in
$A$. Using for $\Phi$ the Z string configuration, it is easy to verify that
the {\it effective} potential becomes
\begin{eqnarray}
V_{eff} \, &=& V_0 + {1 \over 4} <A_i A^i> \bigl( 2 g \sin(\theta_w) \bigr)^2
(\Phi^+)^{\dagger} \Phi^+ \nonumber \\
&=& V_0 + {1 \over 4} \alpha T^2 \bigl( 2 g \sin(\theta_w) \bigr)^2
(\Phi^+)^{\dagger} \Phi^+ \, , \label{poteff}
\end{eqnarray}
where $V_0$ is the bare potential (the last term on the right hand
side of (\ref{stdlag})).

From Eq. (\ref{poteff}) we see that the interaction with the photon plasma
induces a positive contribution to the mass of the charged scalar doublet.
For temperatures in excess of a new critical temperature $T_d$, this
positive contribution will be larger than the negative contribution to the
mass (when expanded about $\Phi = 0$) from the bare potential $V_0$. For
temperatures smaller than $T_d$, the total mass term is negative. From
(\ref{poteff}) and (\ref{stdlag}) we can immediately read off the value
of $T_d$:
\begin{equation} \label{dectemp}
T_d \, = \, \eta \left({2 {\lambda} \over {\alpha}}\right)^{1/2} 
{1 \over {g {\rm sin}(\theta_w)}} \, .
\end{equation}
We thus expect the core of the embedded defect to be symmetric (no
charged scalar fields excited) for $T > T_d$, whereas we expect a
core phase transition\cite{CPT} (charged scalar fields non-vanishing) for
$T < T_d$. However, if we consider the time evolution of an embedded
string which is initially set up with symmetric core (as we will do
in the simulations described below), then the temperature ${\tilde T_d}$
at which the symmetric vortex undergoes a core phase transition is
expected to be lower, since gradient energy is required in order to
produce an asymmetric core, not just the potential energy which enters
the above considerations. 

Let us return to the issue of W-condensate formation and briefly discuss
the stabilization mechanism from this point of view.
The energy gain per unit string length
(computed by integrating from the center of the string to radius $\rho$)
at zero temperature obtained from the formation of a condensate in
which the amplitude of the W-field at the core radius is denoted $W$
is of the order \cite{Perkins}
\begin{equation} \label{engain}
E(\rho)_{\rm gain} \, \sim \, g^2 \eta^2 W^2 \rho^2 
\end{equation}
(this excludes the gradient energy required to form a core condensate).
At finite temperature, there is an energy loss due to the interaction
of the photon with the W-fields. This energy loss was used in \cite{Garriga}
to show that Z strings in strong magnetic fields can be stabilized,
and the corresponding energy per unit string length (integrated
from the center to radius $\rho$) is given by
\begin{equation} \label{enloss}
E(\rho)_{\rm loss} \, \sim \, e^2 \alpha T^2 W^2 \rho^2 \, .
\end{equation}
As is obvious from comparing (\ref{engain}) and (\ref{enloss}), at
sufficiently high temperatures the electroweak string will be stable
towards the formation of a W-condensate. The critical temperature is
of the order $\alpha^{-1/2} \eta$, as in (\ref{dectemp}).

{\bf 3. Core Phase Transition}

We have simulated the evolution of embedded defects in the presence of
a finite temperature charged plasma using a numerical code based on
the one employed in\cite{NB99}. Although gauge fields are not included
in this case, the stability of the string configuration can be established
since it essentially comes from the modification of the effective
potential for the scalar fields and the gauge field should play an
important role only when one string interacts with another one.

First we set up the {\it initial configuration} as a two-dimensional slice
of an infinitely long straight global string which is formed by the
neutral components of the scalar fields and whose center resides in
the midpoint of the lattice.
Although such a highly symmetric configuration might be too ideal,
it would be appropriate to see whether the core phase transition
or the complete decay of the string occurs since the conservation
of the winding number and the scalar field structure at the string core
can be easily checked in contrast to full three-dimensional simulations.
We then add thermal energy to the configuration in the form of kinetic
energy, that is, the time derivative of the scalar fields.
Its amplitude is $0.1 \times T^2$, and the allocation to the four
components of the scalar field is chosen at random. 

Then, the four scalar fields $\phi$ are {\it evolved numerically} on a
two-dimensional lattice by means of the equations of motion derived for
the scalar field with the effective potential (\ref{poteff}).
Neumann boundary conditions are employed.
During each simulation, the background temperature is constant and
the cosmic expansion is not taken into account. However, in order to
reduce the fluctuations of the fields so that it is easier to see
whether the string configuration is preserved or not, an artificial
damping term is introduced. Thus, the evolution equations can be written as
\begin{equation}
\frac{\partial^2\phi}{\partial t^2}-\nabla^2\phi+D\frac{\partial\phi}
{\partial t}=-\frac{\partial V}{\partial \phi}\ ,
\end{equation}
and the numerical value of $D$ is set to be $0.1$ in the simulation.
The inclusion of this damping term and the choice of the numerical value
of $D$ do not affect the essential results concerning the stability
of the string.
All dimensional quantities are rescaled by appropriate powers of the
symmetry breaking scale, $\eta$ and are made to be dimensionless.
Although most simulations are performed in a box size of $1000^2$,
we have checked that the basic results are insensitive to the box size
by executing $3000^2$ and $6000^2$ grid point simulations. The spatial
resolution is $\Delta x = 0.5\eta^{-1}$, and the time steps are chosen as
$\Delta t = {1 \over 10} \Delta x$. The values of the numerical parameters
are chosen as $\lambda = 2.5 \times 10^{-3}$ 
and $\alpha (2g\sin(\theta_w))^2 =2$.
We have calculated various patterns of the temperature and some of
the results are depicted in the figures.

Figures 1 \& 2 show the resulting scalar fields as a function of time
and averaged over various sizes of square boxes centered at the midpoint
of the simulation box, that is, the string core of the initial
configuration, for a high temperature of $T = \eta$.
Figure 1 depicts the amplitude of the neutral scalar field averaged
over boxes of $10^2$ and $50^2$ grid points as well as averaged over
the entire volume ($1000^2$ grid points). 
Figure 2 shows the average of the charged scalar field over the entire
volume (the average values over the smaller volumes are almost identical). 

The width of the field configuration of a Nielsen-Olesen string is about
$\lambda^{-1/2}$. Thus, the averaging in the smallest of our three volumes
(the $10^2$ grid point volume) is probing mainly the core region of
the embedded string, and thus the amplitude of the neutral field component
is small compared to unity, which is equal to the symmetry breaking scale,
$\eta$, in our normalization scheme, whereas it is already of order unity
on the scale of the $50^2$ grid point volume.
The fact that the neutral scalar field averaged over
the smallest box (enclosing the initial electroweak string core region)
remains small indicates that the electroweak string does not decay.
The fact that the charged scalar field averaged over the core region
remains vanishingly small indicates that no core phase transition takes
place: the defect is a symmetric embedded defect.

%%%%%%%%%%
\begin{figure}
\centerline{\psfig{file=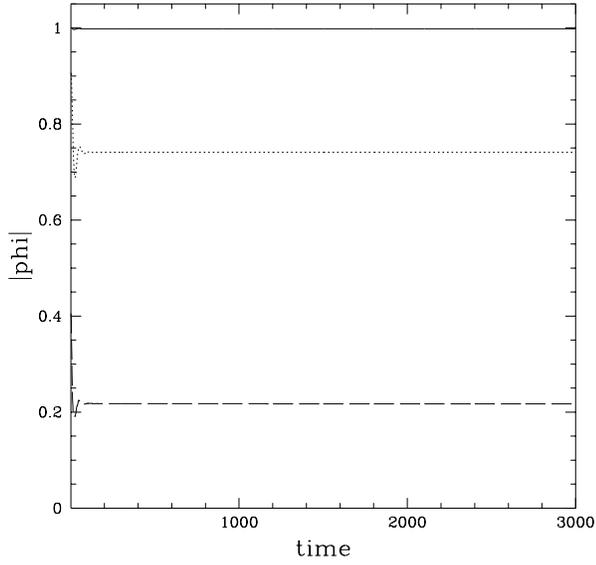,width=8cm}}
\caption{
The amplitude of the neutral scalar field, $\sqrt{\phi_3^2+\phi_4^2}$,
averaged over volumes of $10^2$(dashed line) and $50^2$(dotted line)
grid points, and over the entire volume(solid line), as a function of time,
in the high temperature simulation with $T = \eta$.}
\end{figure}
%%%%%%%%%%

%%%%%%%%%%
\begin{figure}
\centerline{\psfig{file=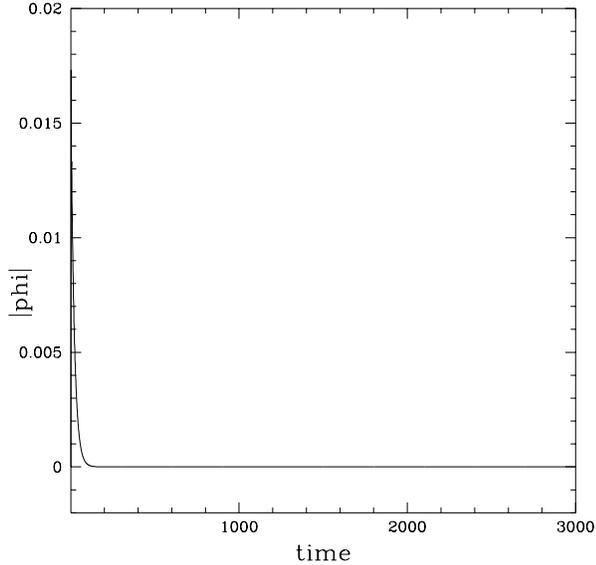,width=8cm}}
\caption{
The amplitude of the charged scalar field, $\sqrt{\phi_1^2+\phi_2^2}$,
averaged over the entire volume, as a function
of time, in the high temperature simulation with $T = \eta$.}
\end{figure}
%%%%%%%%%%

%%%%%%%%%%
\begin{figure}
\centerline{\psfig{file=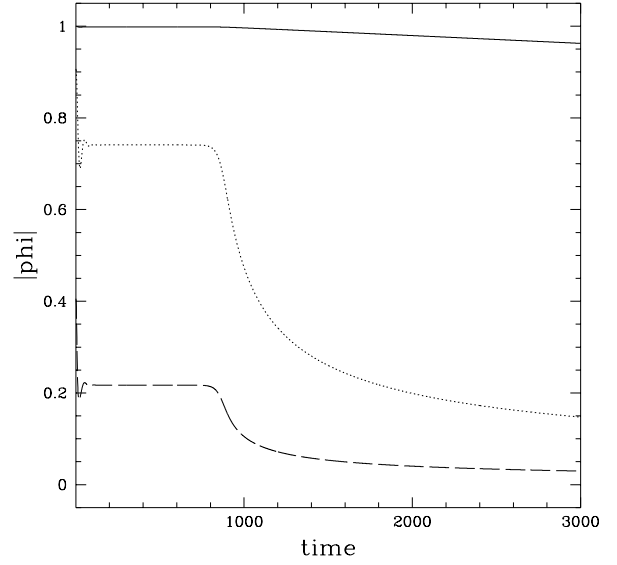,width=8cm}}
\caption{
The amplitude of the neutral scalar field averaged over volumes of
$10^2$(dashed line) and $50^2$(dotted line) grid points, and over
the entire volume(solid line), as a function of time,
in the low temperature simulation with $T = 10^{-3}\eta$. The
core phase transition occurs at a time $t \sim 800$ and is marked by
a sharp decrease in the average value of field in the string core and
surrounding regions.}
\end{figure}
%%%%%%%%%%

%%%%%%%%%%
\begin{figure}
\centerline{\psfig{file=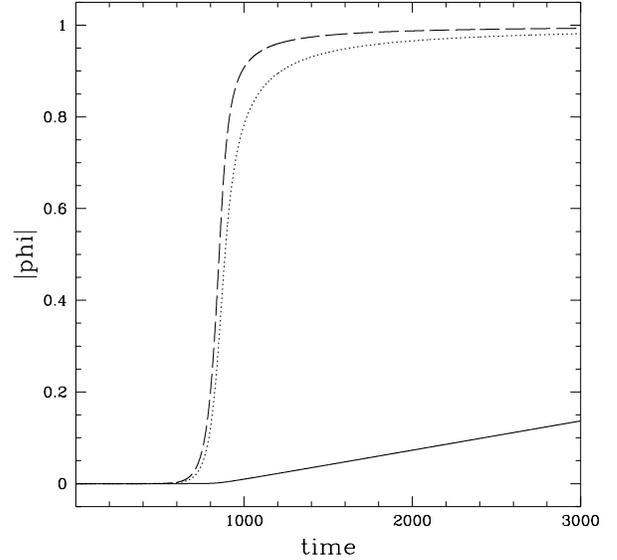,width=8cm}}
\caption{
The amplitude of the charged scalar field averaged over volumes of
$10^2$(dashed line) and $50^2$(dotted line) grid points, and over
the entire volume(solid line), as a function
of time, in the low temperature simulation with $T = 10^{-3}\eta$. At
the time of the core phase transition $t \sim 800$, the charged scalar
field takes on an average value over the string core region and the
neighboring volume of $50^2$ grid points which is of the order $\eta$.}
\end{figure}
%%%%%%%%%%

Figures 3 \& 4 show the corresponding curves in the case of a
much lower temperature $T = 10^{-3} \eta$. Note that this
temperature is lower than the critical temperature $T_d$ when the
effective barrier stabilizing $\phi = 0$ in the effective potential
$V_{\rm eff}$ disappears. The field configuration begins in the same state as
in the high temperature case. However, after a short time a core phase
transition sets in during which the charged scalar field components take
on a non-vanishing value which is of the order of $\eta$. 
The fact that the neutral scalar field
remains small when averaged over the smallest of the three boxes demonstrates
that the embedded string remains stable. 

In fact, the absolute value of
the neutral scalar field components is observed to decrease slightly during
the core phase transition and, in contrast, the amplitude of
the charged scalar fields averaged over the whole simulation box is
increasing. This represents the gradual increase in the core size, an
increase which will stop once the core size $R$ is of the order $T^{-1}$. 
This can be seen as follows: If the core phase transition occurs via
the excitation of a single of the two charged scalar fields (no
winding number in the charged scalar field sector generated), then
the energy is lowered by eliminating the angular gradient energy
of the neutral scalar fields. The energy per unit length thus gained 
will be of the order
\begin{equation}
E(R)_{\rm grad} \sim \eta^2 {\rm ln}\left({R \over w}\right) \, ,
\end{equation}
where $w$ is the core width before the core phase transition. However,
there is an energy cost associated with the generation of a nonvanishing
value for the charged field, and this energy cost (per unit string length)
is proportional to
\begin{equation}
E(R)_{\rm cost} \sim \eta^2 T^2 R^2 \, .
\end{equation}
By balancing the energy gain and the energy loss, one obtains an optimal
core width which is of the order of $R \sim T^{-1}$. Based on this
consideration, we predict that the increase of the average value of
the charged field over the entire box will eventually come to a halt,
on a time scale which is proportional to $T^{-1}$. We have confirmed
these predictions concerning the final value of $R$ and the time scale
in our numerical work.

The time at which the core phase transition takes place
depends sensitively on the value of the damping term added to the
equation of motion; the smaller the damping effect is, the earlier
the transition occurs. However, such a difference is not
cosmologically significant when we consider the stability of
the electroweak string since the time it takes for the core phase
transition to occur is much smaller than the Hubble time
both in either the cases of $T=10^{-3}\eta$ and $T=\eta$.
The asymptotic amplitudes of the fields averaged over the larger two of
the three volumes do not depend on the presence or absence of the damping
term, although the amplitude of the fields in the smallest volume does.
This is due to the fact that in the absence of damping, the thermal
fluctuations will cause the string core to move by more than a core radius.

%%%%%%%%%%
\begin{figure}
\centerline{\psfig{file=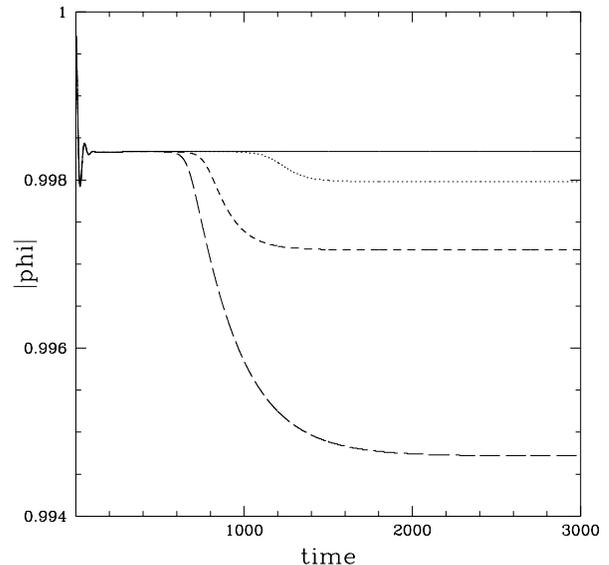,width=8cm}}
\caption{
The amplitude of the neutral scalar field averaged over the entire volume,
as a function of time, for four different temperatures near the critical
temperature ${\tilde T_d}$, $T=0.02\eta$(long dashed line), $T=0.03\eta$
(short dashed line), $T=0.04\eta$(dotted line) and $T=0.05\eta$(solid line).}
\end{figure}
%%%%%%%%%%

%%%%%%%%%%
\begin{figure}
\centerline{\psfig{file=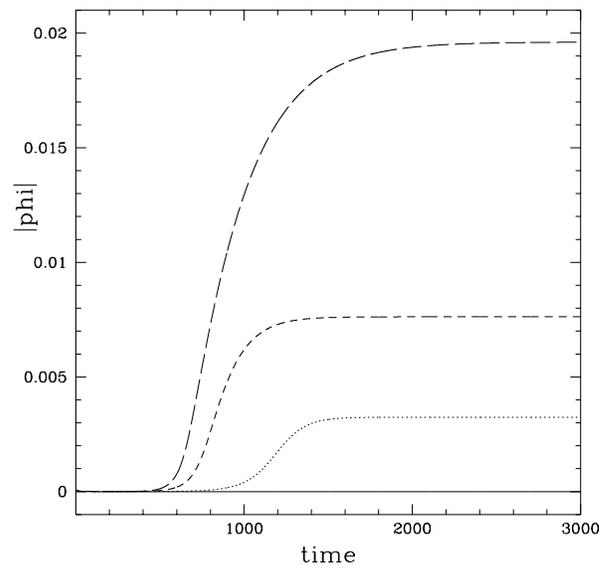,width=8cm}}
\caption{
The amplitude of the charged scalar field averaged over the entire volume,
as a function of time, for four different temperatures near the critical
temperature ${\tilde T_d}$, $T=0.02\eta$(long dashed line), $T=0.03\eta$
(short dashed line), $T=0.04\eta$(dotted line) and $T=0.05\eta$(solid line).}
\end{figure}
%%%%%%%%%%

Note that in both the high and the low temperature simulations, the
winding number (in the neutral scalar field sector) is conserved
not only for the entire simulation box but also when it is calculated
around the string core, the $10^2$ grid points square region.
This is a further test of the stability of the embedded string. 

We have also investigated numerically at which temperature ${\tilde T_d}$ the
core phase transition takes place. We expect \cite{NB99} the phase
transition to happen at the temperature $T_d$
when the potential barrier at $\phi = 0$ due to
the plasma terms disappears. This occurs when the positive quadratic
contribution to $V_{\rm eff}$ due to the plasma equals the negative quadratic
contribution due to the bare potential (see (\ref{dectemp})). 
For the values of $\lambda$
and $\alpha (2g\sin(\theta_w))^2$ chosen the value of $T_d$ is
$0.1 \eta$. However, an initially symmetric core can only undergo
a core phase transition if there is sufficient energy released from
potential energy to create the required gradient energy. Our numerical 
simulations show that this happens at a temperature ${\tilde T_d}$
which lies in the interval between $0.04$ and $0.05$ in units of $\eta$
(see Figures 5 \& 6). Note that the numerical value of ${\tilde T_d}$
for different model parameters such as $\lambda$ or $\alpha$ can be
obtained by simple scaling using (\ref{dectemp}).

Moreover, we have investigated the stability of the embedded electroweak
string at very low temperatures, $T=10^{-6}\eta$.
By tracking the winding number in the neutral field sector,
as well as by tracking the ratio of the neutral to the
charged scalar fields in the string core region, we find no evidence for
a decay of the string. The time evolution of the fields is almost as
shown in Figs. 3 and 4, except for the fact that the core phase
transition sets in later since the thermal fluctuations are weaker. 
Obviously, our approximate analysis is only valid
at temperatures for which the photon is in equilibrium, and thus breaks
down before the cosmic temperature reaches that of last scattering.

{\bf 4. Conclusions and Discussion}

We have studied the stability of the embedded Z-string of the
standard electroweak theory in the presence of an electro-magnetic
plasma in an approximate treatment in which we only follow the
dynamics of the Higgs fields and treat the gauge fields as either
vanishing (the W and Z fields) or else (the photon) as 
being in a thermal bath. We find that the electroweak string is
stabilized at all temperatures above which the photon is in
thermal equilibrium. In the temperature range $T_d < T < T_c$
the embedded string is symmetric in the sense that the charged scalar
fields are not excited in the core, for lower temperatures the
charged scalar fields are non-vanishing in the string core, but
the string itself is preserved in the sense that the winding number in
the neutral scalar field sector is conserved.

The basic mechanism which stabilizes the electroweak Z-string is the
plasma mass for the charged Higgs doublet induced via the
interactions with the plasma. This lifts the vacuum manifold in the
direction of the charged Higgs doublet, leaving an effective
vacuum manifold ${\cal M} = S^1$ which admits stable cosmic string
solutions \footnote{An interesting question is what effect quantum
vacuum fluctuations might have on the possible stabilization of
embedded defects. We are grateful to Leandros Perivolaropoulos
and Bill Unruh for raising this question.}

Our numerical work is based on simulations in which only the
scalar fields are evolved. From a gauge theory perspective
this is an inconsistent approach. However, we feel that since
the stability of the embedded defects is determined by the
scalar field effective potential, neglecting the gauge fields
should not adversely affect our results. However, it would be
interesting to perform full local field theory numerical simulations
to verify our conclusions.

The plasma stabilization mechanism discussed in this work
obviously extends to a wide class of gauge theories which
might be relevant in the early Universe. A new class of
stabilized topological defects will thus create a new and
interesting avenue to explore the interface of particle
physics and cosmology.

{\it Acknowledgments:}

We wish to thank W. Unruh for his kind hospitality at the University of
British Columbia (UBC) where this paper was completed. We are grateful to
 L. Perivolaropoulos, T. Vachaspati and A. Zhitnitsky for useful discussions.
The work of R.B. is supported in part by the U.S. Department of Energy under
Contract DE-FG02-91ER40688, TASK A. R.B. also wishes to thank the
CERN Theory Division and the Institut d'Astrophysique de Paris(IAP) for
hospitality and financial support over the past year.
M. N. wishes to thank IAP for hospitality. He is also grateful
to Kanagawa University for the award of a grant for research abroad.

\end{document}